\documentclass[conference]{IEEEtran}

\usepackage[pdftex]{graphicx}
\usepackage{graphicx}
\usepackage[breaklinks,hidelinks]{hyperref}
\usepackage{balance}
\usepackage{comment}
\usepackage{algorithm}
\usepackage[noend]{algpseudocode}
\usepackage[dvipsnames]{xcolor}
\usepackage{cite}
\usepackage{amsmath}
\usepackage{amssymb}
\usepackage{booktabs}
\usepackage{lipsum}
\usepackage[capitalise,nameinlink]{cleveref}
\usepackage{subcaption}
\usepackage{multirow}
\usepackage{tikz}
\usepackage{caption}
\usepackage{subcaption}
\usepackage{acronym}
\usepackage{enumitem}
\usepackage{siunitx}

\definecolor{darkgreen}{rgb}{0.0, 0.4, 0.0}
\makeatletter
\AtBeginDocument
{
  \def\ltx@label#1{\cref@label{#1}} 
  \def\label@in@display@noarg#1{\cref@old@label@in@display{#1}} 
}
\makeatother

\crefname{equation}{}{}
\graphicspath{ {figures/} }

\usepackage{amsmath} 
\definecolor{ForestGreen}{RGB}{34,139,34}

\title{Automated Testbed for Repeatable Evaluation of Ultra-Wideband Localization Performance}

 \author{\IEEEauthorblockN{Alexander Kemptner\IEEEauthorrefmark{1}\IEEEauthorrefmark{2}, Julian Karoliny\IEEEauthorrefmark{1}, Hannah Brunner\IEEEauthorrefmark{3}, Andreas Gaich\IEEEauthorrefmark{1}, Michael Neubauer\IEEEauthorrefmark{1}, \\Fjolla Ademaj-Berisha\IEEEauthorrefmark{1}, Filippo Casamassima\IEEEauthorrefmark{3},  Walther Pachler\IEEEauthorrefmark{3}, Shrief Rizkalla\IEEEauthorrefmark{1}, Harald Witschnig\IEEEauthorrefmark{3},\\
 Andreas Springer\IEEEauthorrefmark{2}, Hans-Peter Bernhard\IEEEauthorrefmark{1}\IEEEauthorrefmark{2}\\
 \IEEEauthorblockA{\IEEEauthorrefmark{1}Silicon Austria Labs GmbH, Linz, Austria {\footnotesize(alexander.kemptner@silicon-austria.com)}
 \\\IEEEauthorrefmark{2}Johannes Kepler University, Linz, Austria~~~\IEEEauthorrefmark{3}Infineon Technologies Austria AG, Graz, Austria}}}

\begin{document}

\acrodef{wsn}[WSN]{wireless sensor network}
\acrodef{iot}[IoT]{Internet of Things}
\acrodef{nd}[ND]{neighbor discovery}
\acrodef{dc}[DC]{duty-cycle}
\acrodef{rx}[Rx]{receive mode}
\acrodef{tx}[Tx]{transmit mode}
\acrodef{mcu}[MCU]{microcontroller unit}
\acrodef{mac}[MAC]{medium access control}
\acrodef{phy}[PHY]{physical layer}
\acrodef{ble}[BLE]{Bluetooth Low Energy}
\acrodef{ism}[ISM]{industrial, scientific and medical}
\acrodef{cdf}[CDF]{Cumulative Distribution Function}
\acrodef{per}[PER]{packet error rate}
\acrodef{uwb}[UWB]{Ultra-Wideband}
\acrodef{dut}[DuT]{Device-under-Test}
\acrodef{ota}[OTA]{Over-the-Air}
\acrodef{agv}[AGV]{Autonomous Guided Vehicle}
\acrodef{sbc}[SBC]{Single-Board Computer}
\acrodef{ros}[ROS]{Robot Operating System}
\acrodef{ros2}[ROS2]{Robot Operating System Version 2}
\acrodef{dds}[DDS]{Data Distribution Service}
\acrodef{idl}[IDL]{interface description language}
\acrodef{qos}[QoS]{Quality of Service}
\acrodef{ntp}[NTP]{Network Time Protocol}
\acrodef{ptp}[PTP]{Precision Time Protocol}
\acrodef{rest}[REST]{Representational State Transfer}
\acrodef{api}[API]{Application Programming Interface}
\acrodef{gpio}[GPIO]{General Purpose Input/Output}
\acrodef{cep95}[CEP95]{Circular Error Probable 95}
\acrodef{rmse}[RMSE]{Root Mean Squared Error}
\acrodef{dstwr}[DS-TWR]{Double-Sided Two-Way Ranging} 

\maketitle

\begin{abstract}
Testing \ac{uwb} systems is challenging, as multiple devices need to coordinate over lossy links and the systems' behavior is influenced by timing, synchronization, and environmental factors. Traditional testing is often insufficient to capture these complex interactions, highlighting the need for an overarching testbed infrastructure that can manage devices, control the environment, and make measurements and test scenarios repeatable. In this work, we present a highly automated testbed architecture built on \acl{ros2}, integrating device management with environmental control and measurement systems. It includes an optical reference system, a controllable \acl{agv} to position devices within the environment, and time synchronization via \ac{ntp}. The testbed achieves a \acl{rmse} of 4.8\,mm for positioning repeatability and 0.493° for the orientation, and our \ac{ntp}-based synchronization approach achieves a timing accuracy of below 1\,ms. All testbed functionality can be controlled remotely through simple Python scripts to allow automated orchestration tasks such as conducting complex measurement scenarios. We demonstrate this with a measurement campaign on \ac{uwb} localization, showing how it enables repeatable, observable, and fully controlled wireless experiments.
\end{abstract}

\section{Introduction}
\acresetall

Evaluating the performance and firmware of wireless networks for applications like \ac{uwb} localization requires diverse functionality. For example, (i) devices need to be flashed with updated firmware, (ii) power and reset states need to be controllable to correctly initialize and start measurements, and (iii) logs must be collected to assess the localization performance afterwards. Additionally, not only do the devices need to be managed, but the environment should also be controllable and measurable. Therefore, a testbed's functionality should also extend to infrastructure like \acp{agv} and optical reference systems. For \ac{uwb} systems, this infrastructure is crucial to accurately measure localization performance. Unlike most testbeds, we use \ac{ros2} for our implementation. This significantly simplifies the testbed setup, as many infrastructure manufacturers provide pre-existing implementations for \ac{ros2}. Additionally, this allows for a centralized control of the whole testbed through a script or an interactive user interface.

In the literature, numerous testbeds for \ac{uwb} performance measurements have been presented.
One class of testbeds does not directly integrate the \ac{uwb} devices into the testbed architecture, but instead uses a sniffer to record the data~\cite{peterseil2023uwb}. This enables a quick measurement setup for experiments with a focus on data post-processing. However, if many different chips or algorithms need to be tested, the manual orchestration of the devices (flashing firmware, starting/stopping measurements, etc.) becomes time-consuming. \\
More complex, fully integrated testbed setups for \ac{uwb} are shown in~\cite{raza2019indoor},~\cite{delamare2020new}. However, they also lack automation features like firmware updates on the devices, which are critical for quick evaluation of different algorithms. The testbed in~\cite{lcmtestbed} provides this automation in an integrated architecture using the \ac{ros} (Version 1), but does not offer further functionality needed for \ac{uwb} evaluation, like location ground truth measurements. The \textit{Cloves} testbed presented in~\cite{molteni2023cloves} also provides firmware flashing on a large scale, but no integration of \acp{agv} or reference systems. Similarly, the testbed shown in~\cite{schuh2022benchmarking} provides firmware flashing and log collection in a custom architecture, but assumes a fixed position of \ac{uwb} devices, and does not provide live location ground truth measurements. The 5G testbed and measurement campaign reported in~\cite{testbed_damir} includes \ac{uwb} functionality, with the \ac{uwb} subsystem implemented using an early version of the testbed presented in this paper. A fully integrated setup for \ac{uwb} measurements based on \ac{ros2} is used in~\cite{moron2023benchmarking}. However, the authors focus on measurement results and do not go into detail on the setup of the testbed itself. 
In summary, existing literature provides only limited descriptions of fully integrated \ac{uwb} testbed architectures.

In this work, we want to fill this gap by providing a complete description of such a testbed capable of repeatable, automated \ac{uwb} performance measurements. We do not solely focus on the \ac{dut}, but describe all components and their interconnections in detail. We build upon open source implementations which vastly accelerates implementation for readers interested in recreating our setup. In addition, we evaluate key performance aspects of the testbed itself, including its positioning and orientation repeatability as well as its time-synchronization capabilities for both wired and wireless orchestration. Finally, we demonstrate the system through a simple grid-based \ac{uwb} localization example to showcase its end-to-end functionality.\\
The remainder of this work is organized as follows. In \cref{sec:architecture}, we introduce the general architecture of the testbed. In \cref{sec:realization}, the hardware and software realization of this architecture is described in detail, which is the main contribution of this work. \Cref{sec:evaluation} includes the performance evaluation of the testbed and shows results of \ac{uwb} localization measurements using our setup.

\section{Testbed Architecture} \label{sec:architecture}
\begin{figure}
\centering
\includegraphics[]{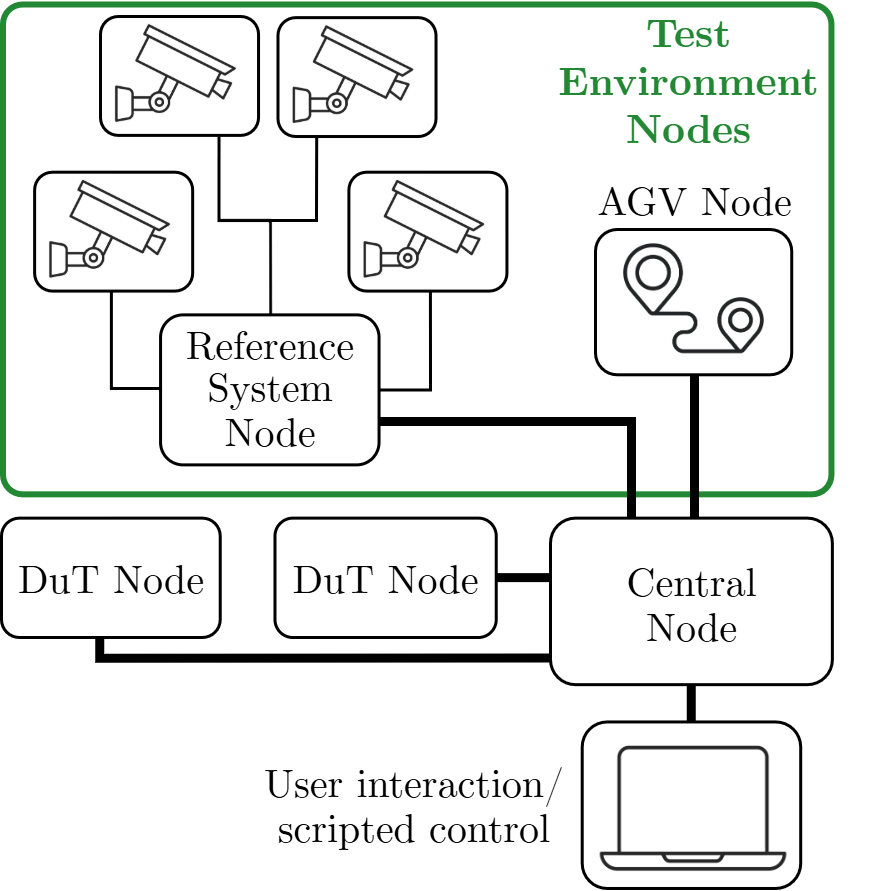}
\caption{Architectural overview of the testbed, its individual building blocks and their connections.} \label{fig:overviewblockdiagram}
\end{figure}

In this section, we highlight the architecture of our proposed testbed, the individual components, and how they are connected. The building blocks of the architecture are referred to as \emph{nodes} in this work. An overview is outlined in \cref{fig:overviewblockdiagram}, where the different node types are depicted. The types can be separated into the \ac{dut} nodes, a central coordination node, and test environment nodes. The test environment nodes control or interact with the environment, such as the reference system node or the \ac{agv} node. Together, the nodes form the architectural base of our fully automated wireless testbed. 

\subsection{DuT Nodes} \label{sec:dutnodes}
The primary component of the testbed is the wireless device with the corresponding firmware that needs to be evaluated, which we refer to as \acf{dut}. In our proposed architecture, it is always paired with a \ac{sbc} as a management device, which provides all testbed orchestration functionality. Together, they form a \textit{\ac{dut} node}. The motivation for this is twofold. Firstly, the firmware of the \ac{dut} is not cluttered with testbed orchestration functionality like \ac{ota} updates. Secondly, the wireless channel used by the \acp{dut} is free from interference, as the management device can use a communication technology different from the \ac{dut}. In the case of our proposed \ac{uwb} testing, the management can be performed via WiFi or Ethernet without interfering with the \ac{uwb} communication. The separation in both firmware and communication channel ensures that measurements only evaluate the \ac{dut} performance, while not being influenced by orchestration tasks. The main functionalities of the \ac{dut} nodes are turning the \ac{dut} on and off, interacting with the firmware, or gathering measurements or logs. 

\subsection{Test Environment Nodes}
Although the \acp{dut} are the main focus of the wireless communication testing, the interaction with the environment is also important. In wireless testing, precise and repeatable measurements are only achievable within a controllable and observable environment. Therefore, in addition to the \ac{dut} nodes, our testbed architecture incorporates \emph{test environment nodes}. This category includes all devices capable of either manipulating or monitoring the environment. Examples of manipulations include altering the positions of environmental objects or adjusting conditions such as lighting and temperature. Conversely, environmental state measurements include activities such as object tracking or monitoring physical parameters. The green box in \cref{fig:overviewblockdiagram} highlights two examples of these node types which are used in our testbed. One of these nodes is the \ac{agv} node. Its purpose is to vary the location of objects in the environment for measurement purposes. As an example, a \ac{dut} node could be moved by the \ac{agv} to assess its behavior under varying spatial configurations. The architecture is flexible with respect to the specific type of \ac{agv}. This could be a ground-based robot, a drone, or another vehicle. The second node is a monitoring node to record the ground truth of locations. In our testbed, this reference system node manages a set of highly precise reference cameras. 

\subsection{Central Node and User Interaction}
The core of the testbed is the central node, which connects all individual nodes as illustrated in~\cref{fig:overviewblockdiagram}. The testbed follows a star topology, with the central node serving as the primary communication hub. It is responsible for maintaining connections with the devices and coordinating various tasks, such as sending control commands to the other nodes and recording measurements.
Another key function of the central node is to provide time synchronization services to all individual nodes, ensuring a common and consistent time base across the system. In addition to managing internal network communication, the central node also serves as the main user interface. Users can interact with the system either by sending commands to the central node interactively or through scripted workflows. The latter allows for automated and repeatable orchestration tasks such as conducting complex measurement scenarios.
Importantly, the user interacts only with the central node. Commands directed to other nodes, e.g., to the \ac{dut} node, are always routed through the central node, which handles the coordination and dispatch of these tasks to the respective devices.

\section{Testbed Realization} \label{sec:realization}
In this section, the realization of the aforementioned testbed architecture is described in detail. In particular, we will focus on the communication between the individual nodes shown in \cref{fig:overviewblockdiagram} and the protocols used to implement this communication. To ensure implementation efficiency and reproducibility, standard and openly available protocols and middleware frameworks were employed. Specifically, \ac{ros2} was chosen for inter-node communication, \ac{ntp} for time synchronization between nodes, and a \ac{rest} \ac{api} for user interaction with the central node. These technologies were selected not only for their technical suitability but also because they are open standards and widely supported. This choice provides a robust foundation for future extensions and facilitates the adoption and replication of the testbed by other researchers.

\subsection{Orchestration with ROS2} \label{sec:orchetrationwithros}
The major part of the communication in the testbed occurs between nodes. For this purpose, \ac{ros2} is used~\cite{ros2022}.
The support for multiple communication patterns, provided software tools, as well as pre-existing integrations from many manufacturers led to this choice.
Originally, \ac{ros2} is a software platform for building robot systems. It provides much of the functionality often needed in these systems and can be extended to fit the specific system with custom code written in C++ or Python. While robot systems are different from a wireless testbed, many of the challenges they face are similar. Both are composed of interconnected nodes, often running on different hosts, which need to work together to achieve a common goal. The interconnection of the nodes requires mechanisms for node discovery and communication, potentially across heterogeneous wired and wireless links as well as different operating systems. \ac{ros2} provides all of these capabilities. Its widespread adoption is further reflected by the many device manufacturers that already provide code to integrate their products into \ac{ros2} systems. This includes \ac{agv} and optical reference system manufacturers.
Other protocols, such as MQTT, as used in the 5G testbed in \cite{testbed_damir}, can achieve comparable orchestration. However, much of the functionality and synergies that \ac{ros2} provides out of the box would have to be implemented manually.

\subsubsection{Networking and Data Exchange}
\begin{figure*}
\centering
\includegraphics[width=182mm]{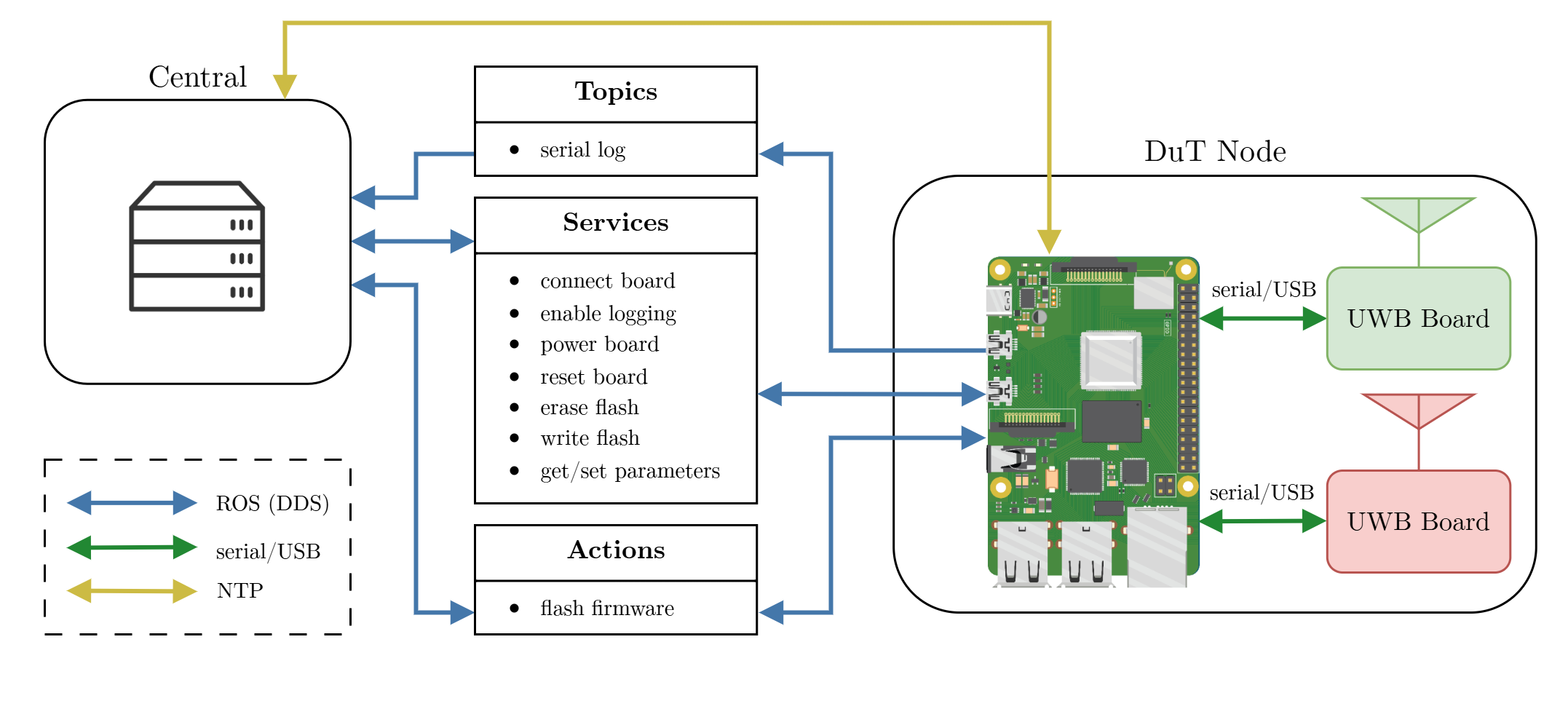}
\caption{Detailed block diagram of information and command flow between the central node and the UWB node.} \label{fig:detailedblockdiagram}
\end{figure*}

In terms of networking, \ac{ros2} serves as a middleware and one of the key features provided is peer-to-peer discovery. This means that nodes do not need to know the IP addresses of other nodes to communicate with them. Rather, they can be addressed by name or enumerated by type (e.g. all \ac{dut} nodes). This is possible even without a central server. Therefore, the central node shown in \cref{fig:overviewblockdiagram} is not a requirement from \ac{ros2}, but a deliberate choice in our architecture to have a central orchestration hub. 
Messages can be exchanged in three different communication schemes. As illustrated in blue in \cref{fig:detailedblockdiagram} for the \ac{dut} node, these schemes are characterized as follows:
\begin{description}
\item[Topics] follow a publish-subscribe pattern. This means that one or many nodes can publish to a specific topic. Nodes can subscribe to this topic and receive the messages.
\item[Services] provide a client-server pattern. The node acting as a client can send a request to a specific node, which answers with a response. The requester waits for the response, so the node is required to answer in a timely manner.
\item[Actions] are a mix of topics and services. A client node sends a request to the server node. The server node starts a long-running task and signals the start via an initial response. Progress can be messaged back to the client node via a topic. When the task finishes, a response is sent.
\end{description}
For both discovery and communication, \ac{ros2} builds upon \ac{dds}, which is an open communication standard. It offers configurable \ac{qos} parameters, which can be used to adjust the communication to available bandwidth, allowed latency, or other requirements. For example, sensor values sent through a topic are not retransmitted if communication errors occur when the \ac{qos} setting \emph{reliability} is set to \emph{best-effort}. Similarly, the application is notified if no values are received for a specified time by setting the \ac{qos} setting \emph{deadline} to this time~\cite{ros2022, rosqos2025}. In addition to its flexibility, studies show good performance of the \ac{dds} protocol in terms of round-trip times~\cite{industry4protocols2019}.

\subsubsection{Example Link -- Central to DuT Nodes}
The communication between a \ac{dut} node and the central node is shown in detail in \cref{fig:detailedblockdiagram}. As explained in \cref{sec:dutnodes}, a \ac{dut} node consists of a management device and one or more \acp{dut}. They are connected via serial over USB, which is indicated by the green arrows. The management device handles all further communication to the testbed. Communication shown with blue arrows is done via \ac{ros2}. All three communication schemes provided by \ac{ros2} are used. The serial log is streamed to the central as a topic which supports multiple nodes subscribing and publishing to it. Services are used for commands controlling the \ac{dut} board. Finally, as flashing the complete firmware is a longer-running task, it is implemented as an action.

\subsection{Time Synchronization Strategy} \label{sec:timesyncstrategy}
In automated testing, reliable measurements are essential, and their validity depends on precise timestamping of the measurements and other relevant events. 
Additionally, accurate synchronization enables faster measurement campaigns, as guard intervals between individual measurements can be kept shorter.
To achieve the highest possible accuracy, timestamping is performed directly at the \ac{dut} node, i.e., at the moment when the \ac{uwb} \ac{dut} sends the information to the \ac{sbc}. The corresponding timing information is then forwarded to the central node via the \ac{ros2} logging topic. Compared to timestamping at the central node, this approach avoids additional latency and jitter introduced by \ac{ros2} communication, resulting in more precise timing data. Since \ac{ros2} itself does not provide time synchronization \cite{rostimesync2021}, the host systems running the \ac{ros2} nodes are synchronized using \ac{ntp}. This is preferred over the more precise \acf{ptp}, as the latter requires all nodes to be connected via Ethernet, and requires special hardware support~\cite{ptprpi52025}, which is not feasible for mobile \ac{agv} and \ac{dut} nodes. After the synchronization procedure, \ac{ros2} uses the clock of the host system for timestamping. In our architecture, the \ac{ntp} server runs on the central node, while all other nodes operate as clients synchronizing to it. This is indicated in \cref{fig:detailedblockdiagram} by the yellow arrow. This setup provides sufficient accuracy for the testbed, even over WiFi, as evaluated in \cref{sec:syncaccuracy}.

\subsection{User Interaction and Automation API} \label{sec:interaction}
All internal communication between the central node and other nodes is handled using \ac{ros2}. The user interacts with the testbed exclusively through the central node. It processes the commands and forwards them as needed. Unlike the internal node communication, the interaction between the user and the central node is based on a \ac{rest} \ac{api}. This approach combines simple interaction with the testbed via \ac{rest}, while maintaining time-critical orchestration via \ac{ros2}. This way, no \ac{ros2} installation is required on the user's device. Additionally, the \ac{rest} \ac{api} offers a well-defined abstraction, which can be used for both interactive and scripted control of the testbed. More specifically, we provide a dedicated interactive web interface as well as a Python API as user interfaces. 

As indicated in \cref{fig:detailedblockdiagram}, the testbed provides extensive functionality to control the \ac{dut} nodes. Users are able to control power to the \ac{dut} devices, which works by switching the USB port on or off. If powered, users have the option to connect the serial port and enable logging. This starts sending logs via the logging topic. Parameters like serial port and baudrate are set and read using the parameter feature provided by \ac{ros2}. Users are able to control the firmware on the board by resetting it, erasing the flash memory, or by flashing a new firmware. 
The latter is implemented as an action, as it is a long-running task. After the user initiates flashing, the central node sends the firmware binary within the \ac{ros2} request message, which the \ac{dut} node confirms with a response. After flashing is finished, the \ac{dut} node signals this to the central node with another response.
As an alternative to writing the complete flash, writing to a specific flash address is also possible, allowing for device-specific configuration such as unique IDs to be set directly in the flash. 

A similar level of control is available for the test environment nodes. For example, the \ac{agv} can be moved to a specific location and orientation and provides internal odometry sensor data, while the reference system node offers access and control of the ground truth location measurements.

\subsection{Software and Hardware Setup}
For the realization of the testbed, we aimed to utilize existing software components as much as possible to allow easy integration of our proposed architecture. This approach highlights one of the key advantages of \ac{ros2}, namely the extensive availability of open-source implementations. For these existing building blocks and nodes, we do not explicitly list the communication endpoints, but instead refer to the respective open-source projects. The reference system node is based on the MOCAP4ROS2 project~\cite{mocap4ros22025}. The \ac{agv} node employs the \ac{ros2} packages provided by the manufacturer, Husarion. For \ac{ntp}-based time synchronization among the nodes, \textit{chrony} is used on both the client side (\ac{dut}, reference system, and \ac{agv} nodes) and the server side (central node)~\cite{chrony2025}.

The main components designed and implemented specifically for the presented testbed are the \ac{dut} and the central node. As shown in~\cref{fig:detailedblockdiagram}, the \ac{dut} node mainly reads and forwards logs from the serial interface and translates received control commands to OpenOCD commands. The central node primarily serves as a bridge between asynchronous user commands and their distribution within the \ac{ros2} domain. As mentioned earlier, this bridging functionality is realized through a \ac{rest} \ac{api}, implemented using the FastAPI framework. The interactive web interface for system control is hosted by the central node and served via the open-source nginx web server.

While the proposed architecture is designed to be flexible with respect to the underlying hardware, the following setup was selected for our realization. The managing part of each \ac{dut} node is realized using Raspberry Pi~5 \acp{sbc}, while the \ac{dut} itself combines a PSOC6 CYBLE-416045 evaluation board with a Quorvo DWM3000EVB \ac{uwb} module. The reference system consists of twelve Prime\textsuperscript{x}~22 cameras by OptiTrack with an accuracy of $\pm\SI{0.15}{\milli\meter}$. For the \ac{agv}, the ROSbot~XL by Husarion is used.
The reference system node and central node are hosted on laptops running Ubuntu 24.04. The software setup on all components is containerized using Docker, including all \ac{ros2} installations as well as the chrony and nginx services. This approach ensures a reproducible and consistent software environment across all devices.

\section{Evaluation: UWB Case Study} \label{sec:evaluation}
In this section, we evaluate the presented functionality of the testbed, including an indoor grid-based \ac{uwb} localization as an example application. All measurements are carried out through the Python scripting interface of the testbed. The \ac{uwb} setup consists of $N=4$ anchors with a fixed and known position on the ceiling in a typical lab environment, and one tag, which is placed on the \ac{agv}. The firmware and configuration for these individual roles are directly flashed onto the corresponding \ac{dut} nodes at the beginning of the measurement scripts by the automated testbed. 
All \ac{dut} nodes operate on \ac{uwb} channel 9 (\SI{7987.2}{\mega\hertz}) at a bandwidth of \SI{499.2}{\mega\hertz}. 

In the first evaluation, we investigate the positioning repeatability achievable through the combination of the \ac{agv} and the optical reference system. Then, we evaluate the synchronization accuracy achievable between \ac{dut} nodes with the described \ac{ntp} solution. Lastly, we demonstrate the testbed end-to-end by executing a simple localization measurement campaign.

\subsection{Positioning Repeatability} \label{sec:repeatability}
To evaluate the positioning repeatability, a fixed reference position and orientation in the room was selected. The \ac{agv} was sent to this position and orientation, followed by a random position and orientation within a grid of $2\textnormal{x}2\textnormal{m}$, then back to the reference position again. This alternating pattern was repeated 100 times. For each position, the real position was measured using the optical reference system and compared to the target. The central node employs a PID controller for moving the \ac{agv} to the defined position within a configurable threshold. Although the controller also uses the very precise reference system as position input, typically there are still differences between the target and real positions due to mechanical constraints of the \ac{agv} and the chosen maximum allowed speed. The thresholds were set to $\pm$\SI{0.5}{\centi\metre}/$\pm$\SI{0.3}{\degree} and the maximum speed to \SI[per-mode=single-symbol]{0.1}{\metre\per\second}. This resulted in a \ac{rmse} for the positioning of approximately \SI{4.8}{\milli\metre}. An additional error measure is the \ac{cep95}, which describes the radius that includes \SI{95}{\percent} of measurements. In our measurements, the CEP95 is approximately \SI{9.32}{\milli\metre}. This proves that overall, the system is capable of repeatable centimeter-level positioning. Similarly, the orientation \ac{rmse} was approximately \SI{0.493}{\degree}, which is especially important when evaluating \acp{dut} with directional antennas. 

\subsection{Synchronization Accuracy} \label{sec:syncaccuracy}

\begin{figure}
\centering
\includegraphics[]{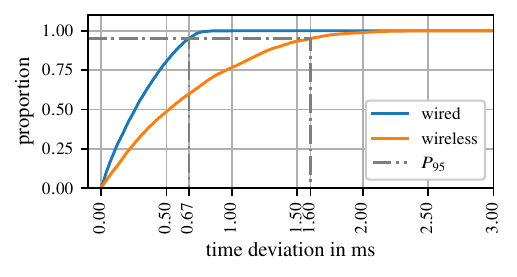}
\caption{\acs{cdf} of NTP synchronization accuracy over wired vs. wireless link.} \label{fig:syncaccuracy}
\end{figure}
To evaluate the \ac{ntp} synchronization accuracy, four \ac{dut} nodes were connected to the central node over a \SI{5}{\giga\hertz} IEEE 802.11ac wireless link. The \ac{dut} node used as the tag was disregarded for this measurement. The \ac{ntp} was configured to dynamically synchronize every $16$ to \SI{256}{\second}. To generate log output, a firmware was flashed onto the \ac{dut}, which writes to the serial interface upon registering a rising edge on a \ac{gpio} pin. As in normal operation of the testbed, this log is collected and timestamped by the \ac{sbc} of the \ac{dut} node. For the measurement, the \ac{dut} nodes could synchronize for 30 minutes. Afterwards, the \ac{gpio} pins were triggered simultaneously at a frequency of \SI{1}{\hertz} for 30 minutes by an arbitrary waveform generator. Ideally, the four timestamps generated on each rising edge by the \ac{dut} nodes should be identical. However, imperfections in the \ac{ntp} synchronization, as well as timing differences in the serial data transmission lead to deviations. These deviations are evaluated by comparing the timestamps for all combinations of two nodes for each rising edge. The results are shown as a \ac{cdf} in \cref{fig:syncaccuracy}. For the wireless connection (orange line), \SI{95}{\percent} of nodes deviate by \SI{1.6}{\milli\second} or less. The \ac{rmse} of time deviations was approximately \SI{0.81}{\milli\second}. As mentioned in \cref{sec:timesyncstrategy}, most test scenarios require at least some nodes to be connected wirelessly for mobility. However, fixed nodes in the testbed are typically connected over Ethernet. The performance for the wired scenario is shown by the blue line in \cref{fig:syncaccuracy}. For \SI{95}{\percent} of nodes, the time deviation is reduced to \SI{0.67}{\milli\second} or less, and the \ac{rmse} was \SI{0.36}{\milli\second}.

\subsection{Testbed-Driven Localization Results} \label{sec:testbedlocalizationresults}
\begin{figure}
\centering
\includegraphics[]{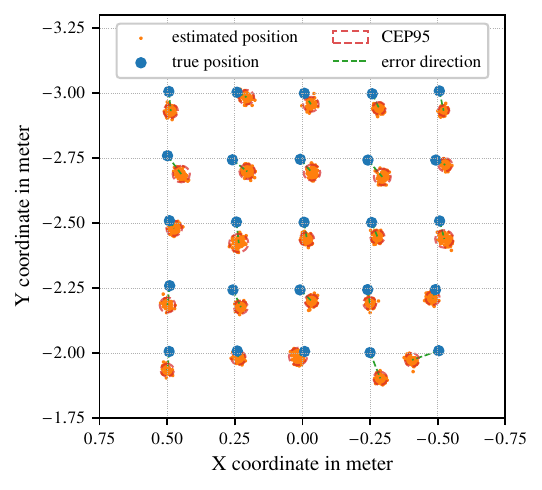}
\caption{View of a measured subgrid.} \label{fig:gridplot}
\end{figure}
To demonstrate the functionality of the testbed end-to-end, a simple measurement campaign was carried out. The \ac{dut} node on the \ac{agv} serving as the tag was moved to positions arranged in a 5x5 \SI{0.25}{\meter} grid pattern. In total, five of these grids were measured at different spots throughout the room in an area of $6 \times 6\,\textnormal{m}$. A zoomed view of one grid is shown in \cref{fig:gridplot}. The true positions measured by the optical reference system and captured by the reference system node are shown by the blue dots. Every time the target position was reached, a reset command was sent to the tag, which initiates 100 \ac{dstwr} \cite{dstwr} exchanges with each of the $N=4$ anchors. This data was collected using the logging functionality of the testbed. Based on these distance measurements $d_i$, we estimate the position $\hat{\mathbf{p}} \in \mathbb{R}^3\ $ of the tag by solving the nonlinear least-squares problem
\begin{align}
    \hat{p} = \arg\min_{\mathbf{p} \in \mathbb{R}^3} \sum_{i=1}^{N} \left( \| \mathbf{a}_i - \mathbf{p} \|_2 - d_i \right)^{2}\,,
\end{align}
where $\mathbf{a}_i$ denotes the position of the $i$-th anchor. We solved this using the Levenberg-Marquardt algorithm. This leads to 100 estimated positions per true position, which are shown as orange dots. The \ac{rmse} of the position accuracy across all grids is \SI{14.56}{\centi\meter}. The red circles show the \ac{cep95} for each position, which visualizes the precision of the position estimation. On average, the radius of these circles is \SI{5.09}{\centi\meter}. The green lines link the true position to the median of the estimated positions. The length of this line visualizes the accuracy of the position estimation. The mean length of these lines is \SI{9.39}{\centi\meter}.

\section{Conclusion and Future Directions}

This work presented a testbed architecture for \ac{uwb} performance measurements. The testbed adopts a flexible architecture covering all aspects needed, like control of \acp{dut}, reference systems, and \acp{agv}. The flexible architecture allows for deployments at new sites with minimal reconfiguration. For internal communication, a hybrid approach was adopted, which uses \ac{ros2} for time-critical internal communication, and a \ac{rest} \ac{api} for the user interface. Both communication protocols have extensive open-source and manufacturer support, which cuts down implementation efforts. As \ac{ros2} does not provide synchronization between nodes, an \ac{ntp} implementation was deployed and tested, which leads to good synchronization errors of below \SI{1.6}{\milli\second} for \SI{95}{\percent} of nodes, also over wireless links. The testbed was evaluated with a simple localization measurement, which demonstrates its end-to-end functionality. Using \ac{dstwr} exchanges, a \ac{rmse} of \SI{14.56}{\centi\meter} was achieved compared to the ground truth provided by the integrated optical reference system. Most importantly, the testbed is capable of automated measurements with high repeatability. By simply executing a Python measurement script again, performance comparisons with updated firmware or different \ac{uwb} chips can be executed quickly. With these features, our testbed will support the development, testing, and ultimately the adoption of many future \ac{uwb} applications like the FiRa Hero Use Cases~\cite{firareport2025} of untracked navigation, public transport fare collection, and asset tracking. 

\bibliographystyle{IEEEtran}
\bibliography{refs}

\end{document}